\documentclass[prl, twocolumn]{revtex4-1}

\usepackage[T1]{fontenc} 
\usepackage{setspace}
\usepackage{lipsum}
\usepackage{graphicx}
\usepackage{printlen}
\usepackage[caption=false]{subfig}
\usepackage{amsmath}
\usepackage{amssymb}
\usepackage{braket}
\usepackage{xcolor}
\usepackage{dsfont} 
\usepackage[retainorgcmds]{IEEEtrantools}
\usepackage{hyperref}

\newcommand{\units}[1]{\ensuremath{ \, \mathrm{#1}}}
\newcommand{\eqnref}[1]{eq.~\ref{#1}}
\newcommand{\figref}[1]{Fig.~\ref{#1}}
\newcommand{\abs}[1]{\ensuremath{\left \vert #1 \right \vert}}
\newcommand{\sgn}{\ensuremath{\mathrm {sgn} }}

\begin{document}

\title{Exact Quantum Electrodynamics in Radiative Photonic Environments}

\author{Ben Yuen}
\email[]{b.yuen@bham.ac.uk}
\author{Angela Demetriadou}
\email[]{a.demetriadou@bham.ac.uk}

\affiliation{School of Physics and Astronomy, University of Birmingham, Edgbaston, Birmingham B15 2TT, United Kingdom}

\date{\today}


\begin{abstract}

We present a comprehensive second quantization scheme for radiative photonic devices. We canonically quantize the continuum of photonic eigenmodes by transforming them into a discrete set of pseudomodes that provide a \textit{complete} and \textit{exact} description of quantum emitters interacting with electromagnetic environments. This method avoids all reservoir approximations, and offers new insights into quantum correlations, accurately capturing all non-Markovian dynamics. This method overcomes challenges in quantizing non-Hermitian systems and is applicable to diverse nanophotonic geometries.
\end{abstract}

\maketitle

The geometry of the environment defines a photon's interaction with matter, bringing complexity to its radiative behaviour; from the Casimir force between an atom and a surface, or the enhanced fluorescence of a molecule by a nanoparticle, to the tapestry of colours scattered by stained glass. 
Lately, quantum emitters (QEs), such as atoms, fluorescent molecules, and quantum dots, have been coupled to ever more geometrically complex photonic devices~\cite{zengin2015realizing, Chikkaraddy2016, benz2016single, sipahigil2016integrated, chikkaraddy2023single, sun2018single}, such as optical microcavities~\cite{Vahala2003}, nanobeams~\cite{Quan2011, sipahigil2016integrated}, plasmonic nanostructures~\cite{zengin2015realizing, Zengin2015, Chikkaraddy2016, benz2016single, chikkaraddy2023single} and hybrid nanophotonic devices~\cite{Doeleman2016,Barreda2022}.  
The interplay with multiple photonic modes of varying radiative behaviour leads to non-Markovian quantum dynamics~\cite{de2017dynamics, breuer2002theory}.
Such dynamics significantly impact the future development of quantum information processing~\cite{flamini2018photonic, wang2020integrated}, quantum transport~\cite{bermudez2013controlling, jezouin2013quantum, pekola2015towards}, photochemistry \cite{mohseni2013geometrical, scholes2017using} and biological processes such as light harvesting~\cite{huelga2013vibrations, hildner2013quantum, pelzer2013dependence}.

The interaction of QEs in photonic environments is usually described by open quantum system coupled to a reservoir that characterises radiative and material loss~\cite{de2017dynamics}. Such descriptions, typical of cavity quantum-electrodynamics (cQED), simplify the continuous electromagnetic spectrum via a simple phenomenologically model of the local density of states and were well suited to early high-finesse micro-resonators and Fabry-Perot cavities that radiate weakly~\cite{yoshie2004vacuum, reithmaier2004strong, peter2005exciton}. 
Lately however, focus has shifted towards ever more intricate photonic devices where  extreme subwavelength field  confinement leads to extreme light-matter interactions~\cite{zengin2015realizing, Chikkaraddy2016, benz2016single, sipahigil2016integrated, chikkaraddy2023single, Quan2011}. Such systems typically radiate efficiently to the far field, exhibit broadband overlapping modes, and often have significant material losses.
Classical models for the local density of states have been phenomenologically quantised to form an open quantum system by assuming Lorentzian photonic modes~\cite{medina2021few, sanchez2022few}.
Alternatively, classical quasi-normal modes~\cite{ge2014quasinormal, lalanne2018light} have been transformed and quantized `canonically'~\cite{franke2019quantization,sauvan2013theory, franke2019quantization, bedingfield2023subradiant}. 
However, such methods lack the generality of established quantum optical theory e.g. \cite{barnett2002methods, lambropoulos2000fundamental},
and their disconnect between the near and far-field precludes straight-forward descriptions of photon statistics, input and output field, squeezed light, dressed states, etc.
Furthermore, reservoir correlations--from which complex non-Markovian dynamics arises--are unaccounted for.
Hence, there is need for an \emph{a priori} method that provides a global picture for open quantum nanophotonic systems.

In this Letter, we develop an \emph{a priori} complete quantum electrodynamic description of light-matter interactions for radiative photonic devices by transforming the quantized fields into pseudomodes~\cite{garraway1997nonperturbative, dalton2001theory}--originally used in an elegant treatment of isolated Lorentzian resonance. 
We show that our generalised pseudomode expansion gives a \emph{complete} and \emph{exact} description of both near and far-field without the need for a reservoir.
The pseudomodes arise naturally from the quantum dynamical equations of motion, yet maintain a direct relationship to the resonant modes of the photonic system.
We derive the pseudomode's quantum equations of motion to obtain \emph{all} the correlations and non-Markovian dynamics of the field, and accurately described the propagation of light.
We give an example application of a QE coupled to a microresonator that supports many spectrally overlapping Mie resonances.

We start with the second quantization of the electromagnetic field, expanded in terms of its eigenmodes $\mathbf u_{\xi}(k,\mathbf {r})$, labelled here by index $\xi$ and wavenumber $k$.
For any nanophotonic system these are the solutions of the Helmholtz equation
\begin{equation}
\left[\nabla^2 + \mu(\mathbf r) \epsilon(\mathbf r) k^2 \right] \mathbf u_{\xi} (k,\mathbf r) = 0,
\label{eq:Helmholtz}
\end{equation}
from which one constructs the corresponding electric and magnetic fields $\mathbf E(k,\mathbf r) = i ck \mathbf u_{\xi}(k,\mathbf r)$, $\mathbf H(k,\mathbf r) = \nabla \times \mathbf u_{\xi}(k,\mathbf r)$.
The nanostructure geometry is specified by the relative electric permittivity $\epsilon(\mathbf r)$--assuming non-magnetic materials ($\mu(\mathbf r)=1$).
To canonically quantise the field via this mode decomposition, the eigenmodes must satisfy the orthogonality condition
$\int_V \mathrm d \mathbf r \epsilon(\mathbf r) \mathbf u_{\xi}(k',\mathbf r) \mathbf u_{\xi'}(k',\mathbf r) =\delta_{\xi \xi'} \delta_{kk'}$
and the energy of each mode must be conserved in time.
For free space ($\epsilon(\mathbf r) = 1$) this is achieved using periodic or zero Dirichlet boundary conditions within a box of volume $V$, which is subsequently taken to infinity.
A similar approach is adopted here, but the shape of the bounding volume $V$ is chosen to match 
the geometry of the nanophotonic device
and the boundary condition $(\mathbf E \times \mathbf B)\cdot \mathrm d \mathbf S=0$ is applied on its surface. The solutions $\mathbf u_{\xi}(k,\mathbf r)$ cover all of space and the discrete spectrum of $k$ becomes continuous when we take the limit $V\to\infty$, which for convenience is performed after quantisation. The quantum field operators are then expanded over $\mathbf u_{\xi}(k,\mathbf r)$, e.g.
\begin{IEEEeqnarray}{rCl}
    \hat{\mathbf E}(\mathbf r) = i\sum_{\xi, \mathbf k} \sqrt{\frac{\hbar c k}{2 \epsilon_0 \epsilon(\mathbf r)}}
    \mathbf u_{\xi}(\mathbf {k,r})
    \left(
     a_{\xi \mathbf k}^{\dagger} +  a_{\xi \mathbf k}
    \right) \label{eq:Eop}
\end{IEEEeqnarray}
where $a^{\dag}_{\xi}(k)$ and $a_{\xi}(k)$ are the usual bosonic creation and annihiliation operators.
QEs interact via the dipole interaction $\hat{\mathbf d} \cdot \hat{\mathbf E}(\mathbf r)$, and the system is described by the Hamiltonian
\begin{IEEEeqnarray*}{rCl}
	H/\hbar &=& \omega_0 \sigma_+ \sigma_- + \sum_{\xi,k} ck a_{\xi}^{\dagger}(k) a_{\xi}(k)\\
	&& + \sum_{\xi ,i, k} g(\mathbf r_i)\left(a_{\xi}(k) \sigma_+^{(i)} + a_{\xi}^{\dagger} (k) \sigma_-^{(i)} \right ). \IEEEyesnumber \label{eq:continuumH}
\end{IEEEeqnarray*}
Each respective term corresponds to the QE with transition frequency $\omega_0$, the field, and their interaction under the dipole and rotating wave approximations, where $\sigma^{\pm}$ are the dipole raising/lowering operators.
The coupling strength for a QE at position $\mathbf r$ is
\begin{equation}
	g_{\xi}(k,\mathbf r) = 
		\sqrt{\hbar c k / 2 \epsilon_0 \epsilon(\mathbf r)} \mathbf d \cdot \mathbf u_{\xi}(k, \mathbf r)
		\label{eq:g_original}
\end{equation} 
where $\mathbf d$ is the dipole moment.

\begin{figure}[t!]
	\includegraphics[width=0.9 \columnwidth]{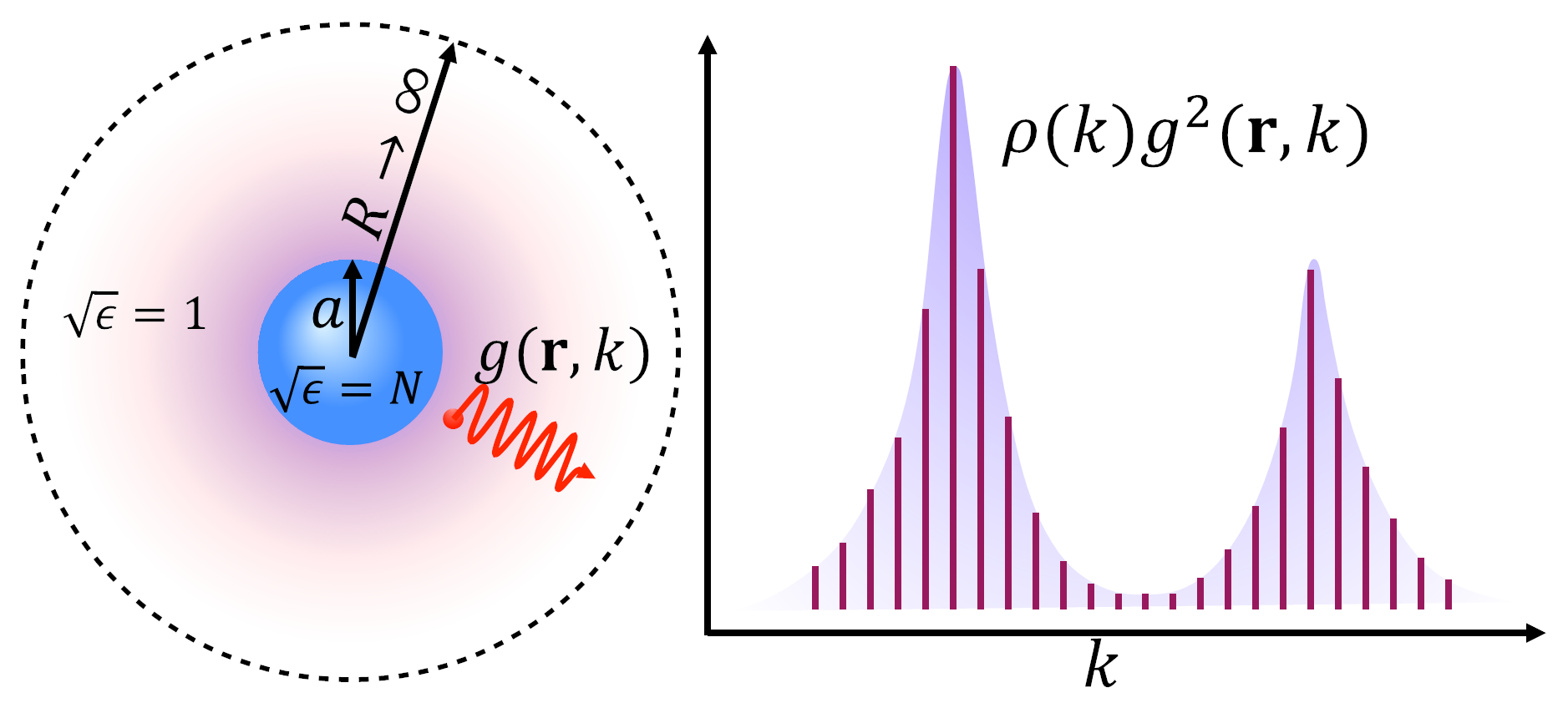}
	\caption{\label{fig:1} 
	Schematic diagram of a spherical nanopartice (blue) surrounded by vacuum and enclosed within a bounding volume (dashed line) of radius $R\rightarrow \infty$. A QE (red) interacts with the mode's electric field (purple). The interaction spectrum (right) is shaped by the nanoparticle geometry.
	}
\end{figure}

We transform the system into the `pseudomode picture' to reveal the full range of quantum dynamics. The mode functions are transformed from the continuous set of Helmholtz solutions $\mathbf u_{\xi}(\mathbf k,r)$  to a discrete set of pseudomodes $\mathbf v_{\xi n}(\mathbf r)$, determined by the integral transformation
\begin{IEEEeqnarray*}{rl}
	\int_{0}^{\infty} k \mathrm d k & \rho(k) \mathbf u_{\xi}(\mathbf {k,r}) \mathbf u_{\xi}(\mathbf {k,r'}) e^{-i c k \tau} = \IEEEyesnumber \label{eq:pseudomode} \\
	&\sum_{n} z_{\xi n} \mathbf v_{\xi n}(\mathbf r) \mathbf v_{\xi n}(\mathbf r') \Theta \left(\tau - \Delta_t(\mathbf r, \mathbf r') \right) e^{-i c z_{\xi n} \tau}
	 ,
\end{IEEEeqnarray*}
evaluated by the residue theorem over poles
$z_{\xi n}$ in the lower half plane for $\tau>0$ (see SI). 
The complex pseudomode frequencies $c z_{\xi n}$ correspond to the $n^{\mathrm{th}}$ resonance of mode $\xi$. 
The integrand's asymptotic behaviour determines the pole location and half-plane the contour integral encloses, and leads to the Heaviside-step function $\Theta \left ( \tau - \Delta_t(\mathbf r, \mathbf r') \right)$. Physically, this accounts for the finite time delay for light to propagate from $\mathbf r'$ to $\mathbf r$ via the photonic structure.

The pseudomode transformation arises naturally from the Schr\"odinger equations of motion with the QE initially excited--we consider general initial conditions using the Heisenberg equations elsewhere \cite{yuenother}. Here the state vector is described by the amplitudes $c_0(t)$ and $c_{\xi k}(t)$ of the QE excited state $\ket{0,e}$ and singly excited field mode $a^{\dagger}_{\xi k}\ket{0,g}$ respectively. 
These obey the coupled equations $\dot {\tilde c}_0(t)=-i\sum_{\xi k} g_{\xi k} e^{i(\omega_0-ck)t}\tilde c_{\xi k}(t)$ and $\dot {\tilde c}_{\xi k}(t)=g_{\xi k}e^{-i(\omega_0-ck)t}\tilde c_0(t)$, derived from the Schr\"odinger equation in the interaction picture.
Formally integrating the later with $\tilde c_{\xi k}(0)=0$, inserting into the former, and taking the continuum limit $V\to\infty $, where $\sum_k \to \int \rho(k) \mathrm d k$ gives
\begin{IEEEeqnarray}{rCl}
	\frac{\mathrm d}{\mathrm d t} \tilde c_0(t) &=& 
		- \int_0^t \mathrm d t' \sum_{\xi} 
		\int \mathrm d k \rho(k) g_{\xi}^2(k) e^{i(\omega_0 - ck)(t-t')} \tilde c_0(t') \label{eq:c0_eom}.
		\IEEEeqnarraynumspace
\end{IEEEeqnarray}
The first Markov approximation $g_{\xi}(k) \approx \sqrt{\gamma / 2 \pi}$ \cite{gardiner1985input} is often applied here to simplify the integral over $k$.
Instead we evaluate the $k$-integral exactly using $g_{\xi}(k,r)$ of \eqnref{eq:g_original} and the transformation of \eqnref{eq:pseudomode} to obtain
\begin{IEEEeqnarray}{rCl}
	\frac{\mathrm d}{\mathrm d t} \tilde c_0(t) &=& 
		- \int_0^t \mathrm d t' \sum_{\xi, n} 
		\bar g_{\xi n}^2 e^{i(\omega_0 - c z_{\xi n})(t-t')} \tilde c_0(t') \label{eq:c0_pm_eom},
		\IEEEeqnarraynumspace
\end{IEEEeqnarray}
This is equivalent to the discrete set of coupled pseudomode equations
\begin{IEEEeqnarray}{rCl} \label{eq:pseudomode_Seom}
		i\frac{\mathrm d}{\mathrm d t} \tilde c_0(t) &=& 
		\sum_{\xi n} \bar g_{\xi n} e^{i(\omega_0- cz_{\xi n})t} \tilde b_{\xi n}(t),
	\label{eq:c0pseudoschrod} \IEEEyesnumber \IEEEyessubnumber \\ 
	i\frac{\mathrm d}{\mathrm d t} \tilde b_{\xi n}(t) &=& 
		\bar g_{\xi n} e^{-i(\omega_0- cz_{\xi n})t} \tilde c_0(t).
	\label{eq:bnpseudoschrod} \IEEEyessubnumber
\end{IEEEeqnarray}
with pseudomode-QE interaction strength
\begin{equation}
 \bar g_{\xi n}(\mathbf r) = \sqrt{\frac{\hbar c z_{\xi n}}{2 \epsilon_0 \epsilon(\mathbf r)}} \mathbf d \cdot  \mathbf v_{\xi n}(\mathbf r). \label{eq:gbar}
\end{equation}
By solving these equations for an initially excited QE one obtains the full quantum dynamical evolution of the system.
These equations are non-Hermitian since $z_{\xi n}$ and $\bar g_{\xi n}$ are complex-valued, which cause the psuedomode amplitudes $\tilde b_{\xi n}(t)$ to decay in time as energy is radiated to the far field.

\begin{figure}[t!]
	\includegraphics[width=\columnwidth]{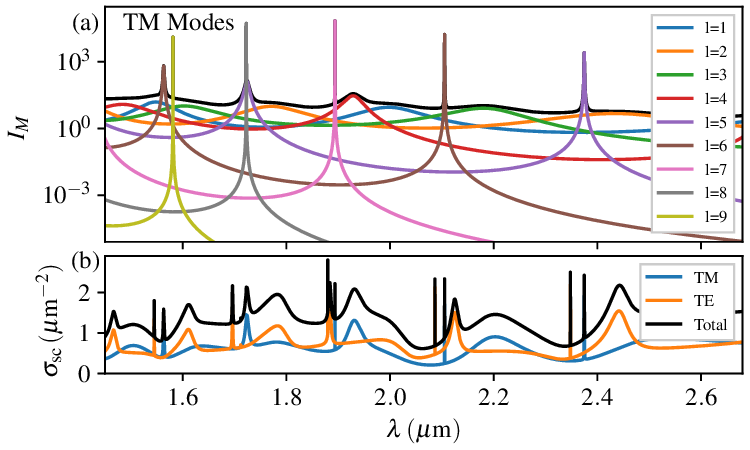}
	\caption{\label{fig:2} 
	Spectrum of a $1\units{\mu m}$ Silicon sphere. (a) The enhancement of TM modes given by the mode normalisation $I_M(k)$ (\eqnref{eq:IM}), with the black line showing the sum over all modes (total) and (b) the scattering cross-section.
	}
\end{figure}

The electromagnetic fields are expressed exactly by a time-dependent superposition of pseudomodes $\mathbf v_{\xi n}(\mathbf r)$. The intensity $\langle E_{-}(\mathbf r,t) E_{+}(\mathbf r,t) \rangle$, is found from the field quadrature acting on $\ket{\psi(t)}$,
\begin{IEEEeqnarray}{rCl} \label{eq:Eplus_op}
	\mathbf E_{+} \ket{\psi(t)} &=& 
		i \sum_{\xi, n} \sqrt{\frac{\hbar c z_{\xi n}}{2 \epsilon_0 \epsilon(\mathbf r)}} 
		\mathbf v_{\xi n} (\mathbf r)
		\tilde b_{\xi n}(t-\Delta t) e^{-i c z_{\xi n}t}, \IEEEeqnarraynumspace
\end{IEEEeqnarray}
which was derived using the pseudomode transformation (see SI).
Here,  $\tilde b_{\xi n}$ is taken at the retarded time $t-\Delta_t(\mathbf r, \mathbf r')$. Expanding the exponential as $e^{-icz_{\xi n}(t-\Delta_t)}e^{-icz_{\xi n}\Delta_t}$ highlights the decay by the factor $e^{-i c \mathfrak{Im}(z_{\xi n}) \Delta_t}$ due to the propagation delay $\Delta_t(\mathbf r, \mathbf r')$. This decay ensures the field remains regular even though the pseudomodes themselves diverge as $r\to\infty$.
Hence our pseudomode approach overcomes the mode divergence problem for radiating photonic systems~\cite{sauvan2013theory, franke2019quantization} to give a global description of the field.

To comprehensively demonstrate this new approach, we apply it to a QE coupled to a spherical silicon resonator of radius $a=1\units{\mu m}$ and refractive index $N=3.446$ surrounded by vacuum (see \figref{fig:1}). We use dimensionless units $\hbar=c=1$ hereafter.
This geometry supports Mie resonances with the near field enhanced due to the confinement of modes by the dielectric interface at $r=a$~\cite{bohren2008absorption}. 
For microspheres with $a \gg \lambda$ the Mie modes interfere constructively, forming high-finesse whispering gallery modes~\cite{oraevsky2002whispering}. 
For $a \lesssim 1\units{\mu m}$, individual Mie modes form a spectrum of distinct but overlapping resonances (e.g. \figref{fig:2}).
The eigenmodes take the form of vector spherical Harmonics 
$ \mathbf M_{lm}(k,\mathbf r) = \boldsymbol {\nabla} \times \hat {\mathbf r} \psi_{lm}(k,\mathbf r) 
$ and $
\mathbf N_{lm}(k,\mathbf r) = \frac1k \boldsymbol \nabla \times \mathbf M_{lm}(k,\mathbf r)$ where $\psi_{lm}(k,\mathbf r) = e^{im\phi} P_l^m (\cos \theta) Z_l(k r)$.
The angular distributions for different integers $l\in \mathds N$ and $\vert m \vert \leq l$ are orthogonal and $P_l^m(\cos \theta)$ are the associated Legendre polynomials (see SI).
The radial functions $Z_l(kr)$,
\begin{equation}
	Z_l(\sqrt{\epsilon} k r) =
        \frac{1}{\sqrt{I_M(k)}} \left\{
            \begin{array}{lr}
                \eta_l(k)
                j_l(\sqrt{\epsilon}kr) \ & r< a \\
                \alpha_l(k)j_l(kr) + \beta_l(k)y_l(kr), \ &r>a
            \end{array}
        \right. , \label{eq:zl}
\end{equation}
are piecewise continuous solutions of the spherical Bessel equation that are regular at the origin, zero on the surface of the bounding volume of radius $R\gg a$, and produce a continuous tangential field at the dielectric interface. 
The interface conditions determine the coefficients $\alpha_l(k), \ \beta_l(k)$ and $\eta_l(k)$ (see SI).
These produce the resonances in \figref{fig:2}, determined by the normalisation factor $I_M(k)$, 
\begin{equation}
	\lim_{R\to\infty} I_M(k)=\frac{R}{2 k^2} (\alpha_l(k)+i \beta_l(k))(\alpha_l(k)-i \beta_l(k)), \label{eq:IM}
\end{equation}
analogous to the mode volume of the photonic system. Resonances occur for real $k$ adjacent to the complex roots $z_{ln}$ of \eqnref{eq:IM} which are poles of \eqnref{eq:zl}.
Similarly, the scattering cross-section $\sigma_s(k)= k^{-2}\sum_l (2 l+1) \vert \beta_l / (\alpha_l+i\beta_l) \vert^2$ shown in \figref{fig:2}b is resonant at the same poles, which have long been identified as the natural modes~\cite{stratton2007electromagnetic}.
Furthermore, the radial functions $Z_{l}(kr)$ are orthogonal due to the conditions at $r=0,a$ and $R$, and therefore $\mathbf M_{lm}(k,\mathbf r)$ and $\mathbf N_{lm}(k,\mathbf r)$ satisfy the orthonormality condition of $\mathbf u_{\xi}(k,\mathbf r)$ with $\xi = (l,m,n)$. 
Finally, for finite $R$, the allowed values of $k$ for which $Z_l(kR)=0$ form a countably infinite set, with asymptotic mode density $\rho(k) = R/\pi$. In the limit $R \to \infty$, $k$ becomes continuous.

\begin{figure}[t!]
	\includegraphics[width=\columnwidth]{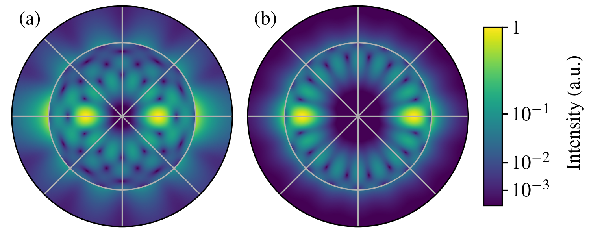}
	\caption{\label{fig:3} 
	Intensity distribution for (a) the $(5,0,4)$ and (b) the $(8,0,3)$ pseudomodes plotted as function of $0<r<1.5\units{\mu m}$ and $\theta$. The resonator surface is at $r=1$.
	}
\end{figure}

\begin{figure}[t!]
	\includegraphics[width=\columnwidth]{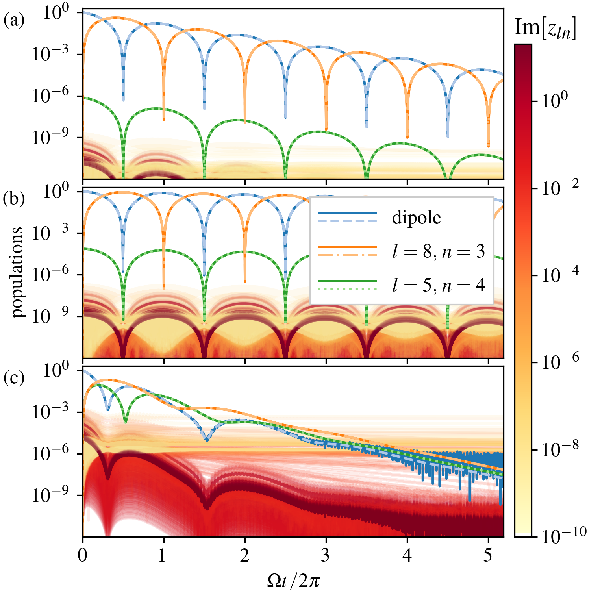}
	\caption{\label{fig:4} 
	Dynamical evolution of QE and pseudomodes state populations calculated for dipole moments of (a) $10\units{Db}$, (b) $100\units{dB}$ and (c) $10^4\units{Db}$. The QE excited state (blue), $(8,3)$ pseudomode (orange) and $(5,4)$ pseudomode (green) populations. The lines in red-yellow hue show all other psuedomode populations where the hue indicates their decay rate given by $\mathfrak{Im}(z_{ln})$. The dashed lines show the approximate results of \eqnref{eq:pseudoEOMapprox}.
	}
\end{figure}

We now quantise these eigenmodes and subsequently perform the pseudomode transformation.
The second quantised electric field operator \eqnref{eq:Eop} is now obtained using these orthonormal modes, and the Hamiltonian is given by \eqnref{eq:continuumH} with coupling strengths given by \eqnref{eq:g_original}.
For simplicity we choose a QE radially oriented $\mathbf d=\vert \mathbf d \vert  \hat{\mathbf r}$ which only couples to the TM modes, and therefore $\mathbf u_{lm}(k,\mathbf r) = \mathbf N_{lm}(k,\mathbf r)$ is sufficient henceforth.
We now transform the system into discrete set of pseudomodes $\mathbf{v}_{lm n}(\mathbf{r})$ that interact with the QE, by evaluating the integral of \eqnref{eq:pseudomode} over $\mathbf N_{lm}(k,\mathbf r)$.
Separating out the $k$-dependent terms, this becomes 
$\boldsymbol{\mathcal N}_{lm}(\mathbf r) \boldsymbol{\mathcal{N}}_{lm}^T(\mathbf r') I_{l}(r,r')$ where
\begin{IEEEeqnarray}{rl}
	I_l(r,r') =& 
		\int_{0}^{\infty} \mathrm d k \rho(k) k^{-1} 
		Z_l(\sqrt{\epsilon} kr) Z_l(\sqrt{\epsilon} k r') e^{-ic k \tau},
		\IEEEeqnarraynumspace
		\label{eq:pseudoint2}
\end{IEEEeqnarray}
and $\boldsymbol{\mathcal N}_{lm}(\mathbf r) \boldsymbol{\mathcal{N}}_{lm}^T(\mathbf r')$ is the outer product of the differential operators such that $\mathbf N_{lm}(k,\mathbf r) = \boldsymbol{\mathcal N}_{lm}(\mathbf r) Z_l(\sqrt{\epsilon} k r) / k$ (see SI).

To evaluate \eqnref{eq:pseudoint2}, initially with lower limit $k=-\infty$, we extend the integrand $f(k)$ into the complex plane $k\to z$.
We expand $Z_l(\sqrt {\epsilon} z r)$ into its outgoing and incoming components proportional to $h_l^{(^1_2)} = j_l \pm i y_l$.
This splits $f(z)$ into incoming $f_1(z)$, outgoing $f_2(z)$ and mixed $f_{12}(z)$ components that converge to zero as $\abs{z}\to \infty$ in the lower half plane when $c \tau > (r-a)$ for $f_1$, $c \tau < - (r-a)$ for $f_2$, and $c\tau < r$ for $f_{12}$, and the upper half plane otherwise.
After careful consideration of the poles and contours (see SI), we find
\begin{IEEEeqnarray*}{rCl}
	I_l &=& 
		\Theta\left[c\tau-(r-a)\right] \oint_{\mathrm{LHP}} f_1(z) \mathrm d z \\
		&&+\Theta\left[c\tau+(r-a)\right] \oint_{\mathrm{UHP}} f_2(z) \mathrm d z,
		\IEEEyesnumber
		\\
\end{IEEEeqnarray*}
which is evaluated using the residue theorem. To evaluate \eqnref{eq:pseudoint2} with lower limit $k=0$, we split $f(k)$ further, into symmetric $f_s(k)$ and antisymmetric components $f_a(k)$, then integrate $f_s(k)$ or $\sgn(k )f_s(k)$ as before then divide the result by two (see SI).
Consequently, for positive times,
\begin{IEEEeqnarray}{rCl}
	I_l(r,r') = 2 \pi i \sum_{z_{ln} \in \mathrm{Q_4}} \mathrm {Res} \left[ f_2(z) \right] \Theta \left( c\tau - (r-a) \right)
\end{IEEEeqnarray}
where the sum is over the fourth quadrant of the complex plane.
\begin{IEEEeqnarray}{rCl}
	\mathbf v_{lmn}(\mathbf r) &=& \boldsymbol{\mathcal N}_{lm}(\mathbf r) \left[ \pi
        \sqrt{ 
		\frac{\alpha_l(z_{ln}) - i \beta_l(z_{ln})}{i \left[ \partial_z (\alpha_l(z) + i \beta_l(z))\right]_{z_{ln}}}
		}
	h_l^{(1)}(z_{ln} r) \right]. \IEEEeqnarraynumspace \label{eq:zbar}
\end{IEEEeqnarray}
where the term in brackets defines the pseudomodes radial behaviour.
These pseudomodes differ from $\mathbf u_{lm}(k,\mathbf r)$ in their radial behaviour, complex nature, and their discrete spectrum over the roots of \eqnref{eq:IM} in the fourth quadrant.
We show the distributions of two different pseudomodes, $(l,m,n)$ equal to $(5,0,4)$ and $(8,0,3)$ in \figref{fig:3}. These correspond to the two narrow overlapping resonances at $1.72 \mu m$ in \figref{fig:2}a.

The quantum dynamics of the system are found using \eqnref{eq:pseudomode_Seom}. 
Choosing the QE coordinate $\theta=0$ simplifies further calculations since $Y_l^m(0,\phi)\equiv0$ for $m\neq0$, an we label modes by (l,n) henceforth.
We set the QE's shifted transition frequency 
($\omega_0+\mathfrak{Im}[\delta \omega_0]$) to be resonant with the $(8,3)$ pseudomode, where
\begin{equation}
	\delta \omega_0 =
		\sum_{\mathrm{off-res.}} g_{ln}^2
		\frac{\left(\omega_0 - \omega_{ln}\right) - i \gamma_{ln}}
			{(\omega_0 - \omega_{ln})^2+\gamma_{ln}^2}, \label{eq:deltaomega0}
\end{equation}
is the Lamb shift due to off-resonant modes (see SI).
The dynamical equations for the $613$ psuedomodes in the range $l\leq 30$ and $\mathfrak {Re} z_{ln} \vert < 20 \units{\mu m}$ are solved rapidly via a similarity transformation that diagonalises \eqnref{eq:pseudomode_Seom}.

Figure \ref{fig:4} shows the time evolution of the QE excited state (blue lines) and pseudomode populations. 
For $d=10\units{D}$, Rabi oscillations (inherently non-Markovian by nature) occur between the QE and the strongly coupled (8,3) mode at frequency $\abs{2 \bar g_{8,3}}$, 
and their decay is primarily due to the radiation of the $(8,3)$ mode. By contrast, the weakly coupled $(5,4)$ mode dynamics are Markovian, oscillating in-phase with the QE six orders of magnitude lower. 
All other modes are shown by the background (yellow-red lines) with small populations $\lesssim 10^{-9}$, and are initially rapidly excited to values $\sim \abs{g / (\omega_0 - c z_{ln})}^2$ (see SI).
For $d=100\units{D}$ in \figref{fig:4}b, the Rabi oscillations are 10 times faster, whilst the decay rate remains largely unchanged. 
For $d=10^4\units{D}$ (\figref{fig:4}c) the dynamics are more complex. 
The (5,4) more, on the cusp of strong coupling, now oscillates out of phase with the QE and radiates energy more efficiently than the $(8,3)$ mode, significantly accelerating the QE's decay.
The remaining non-resonant mode's behaviour depends on the ratio $\gamma_{ln}/\Gamma_0$, where the rate $\Gamma_0$ characterises the QE decay:
Strongly radiating pseudomodes ($\gamma_{ln}/\Gamma_0>>1$, red lines) follow the QE amplitude with $b_{ln}(t) \approx g_{ln} c_0(t) / (\omega_0 - c z_{ln})$ demonstrating Markovian behaviour.
High-Q pseudomodes ($\gamma_{ln}/\Gamma_0<<1$, yellow) are non-Markovian, evolving as $b_{ln}(t) \approx g_{ln} \exp(-i c z_{ln} t) / (\omega_0 - c z_{ln})$ for long times $\Gamma_0^{-1} t \gg 1$.

We can approximate our system, dominated by the $(8,3)$ and $(5,4)$ modes, by a two-mode model given by
\begin{IEEEeqnarray}{rCl}
	i \frac{\mathrm d}{\mathrm d t} \tilde c_0(t) &=& 
		\delta \omega_0 \tilde c_0(t) +
		\sum_{^{(5,4),}_{(8,3)}} \bar g_{ln} e^{i(\omega_0-cz_{ln})t}\tilde b_{ln}(t),
		\IEEEeqnarraynumspace \label{eq:pseudoEOMapprox}
\end{IEEEeqnarray}
and \eqnref{eq:bnpseudoschrod} for these two modes only, whilst $\delta\omega_0$ is calculated from \eqnref{eq:deltaomega0} over all other modes.
These approximate solutions are plotted in \figref{fig:4} with dashed lines and show remarkable accuracy provided $\abs{c_0}^2$ is larger than the off-resonant mode populations.

\begin{figure}[t!]
	\includegraphics[width=\columnwidth]{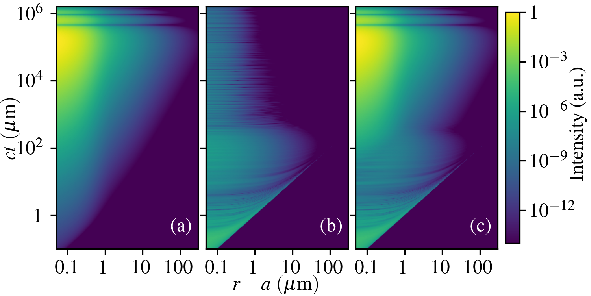}
	\caption{\label{fig:5} 
	Expected field intensity $\langle I(\mathbf r,t)\rangle$ for a $d=10\units{D}$ QE as a function of radius $r$ and time $c t$ for (a) the (8,3) pseudomode, which is localised around the photonic resonator, (b)  all other modes showing radiative behaviour and (c) total intensity.
	}
\end{figure}

Finally, we calculate the expected field intensity $\langle I(\mathbf r,t) \rangle =  \langle \mathbf E_-(\mathbf r,t) \mathbf E_+(\mathbf r,t) \rangle$ using \eqnref{eq:Eplus_op} for $d=10\units{D}$ (see SI for animations). Figure \ref{fig:5} shows $\langle I(\mathbf r,t) \rangle$ plotted against distance from the sphere ($r-a$), and time ($c t$), and clearly shows the light cone.
The contribution from the high-Q $(8,3)$ mode (\figref{fig:5}a), confined to the near field $r-a\lesssim 1\units{\mu m}$, shows decaying Rabi oscillations with period $cT = 4.63 \times 10^5 \units{\mu}$. 
All other modes are shown in \figref{fig:5}b. 
A short pulse originates in the region $(r=a, \ ct \lesssim 1\units{\mu m})$, 
due to transient excitation of the field.
Radiation from the weakly coupled $(5,4)$ mode then dominates in the region $1\units{\mu m} <c t< 200\units{\mu m}$, and interferes with more weakly excited modes. 
For $c t >200\units{\mu m}$, the $(5,4)$ mode has radiatively decayed, revealing non-Markovian dynamics in the near field due to high-finesse off-resonant modes.
Figure \ref{fig:5}c shows the total intensity is dominated at short times by the off-resonant modes, and at longer times by the coherent oscillations of the $(8,3)$ mode.

By transforming the continuum into a discrete set of pseudomodes, we solve the dynamics without the need for a reservoir, its accompanying approximations,  
and therefore we retain all the information about the continuum.
Our theory demonstrates that QE decay arises from the continuous nature of the quantised field:
The system never decays to the ground state $\ket{0,g}$; its
energy is merely dispersed over the continuous spectrum of eigenmodes, for which there is no inherent decay. 
The`decay' of the pseudomode amplitudes $\tilde b_{\xi n}(\tau)$ describe tail of the photon wavepacket as it propagates, when taken correctly at the retarded time $\tau=t-r/c$. This arises naturally in our infinite, yet closed quantum system, which avoids outgoing wave boundary conditions on the normal modes $\mathbf u_{\xi}(k,r)$ that necessitate non-standard quantization of divergent modes for non-Hermitian systems~\cite{franke2019quantization}.

In conclusion, we present a general theory that gives a \emph{complete} and \emph{exact} description for the quantum electrodynamics of a QE strongly interacting with a radiating photonic device.  
We quantise the continuous Helmholtz eigenmodes, which we then transform with one-to-one correspondence into a discrete set of non-Hermitian pseudomodes. 
Thus, we solve common problems met when quantisating non-Hermitian systems, such as mode divergence, defining mode volumes, and identifying canonical field variables.
Furthermore, our approach precisely captures all quantum correlations of the field and QE, avoiding common Markovian approximations, and unlike other methods, accurately captures the light propagation to the far-field.  
This new method can be further extended for arbitrary photonic geometries through the analytic continuation of the local density of states, and can reveal the non-Markovian behaviour exhibited in experimentally realisable systems at the nanoscale.

\vspace{0.5cm}
\begin{acknowledgments}
AD gratefully acknowledges support from the Royal Society University Research Fellowship URF/R1/180097 and URF/R/231024, Royal Society Research Fellows Enhancement Award RGF /EA/181038, and funding from EPSRC EP/X012689/1, EP/Y008774/1 and CDT in Topological Design EP/S02297X/1.
\end{acknowledgments}

\bibliography{paper}

\end{document}